# Advancing AI Audits for Enhanced AI Governance

Arisa Ema [1], Ryo Sato [2,3], Tomoharu Hase [4,5], Masafumi Nakano [6,7*], Shinji Kamimura [8], Hiromu Kitamura [9,10]

[1] Associate Professor, Institute for Future Initiatives, The University of Tokyo
[2] Visiting Researcher, Institute for Future Initiatives, The University of Tokyo
[3] Deloitte, Touche Tohmatsu LLC
[4] Visiting Researcher, Institute for Future Initiatives, The University of Tokyo
[5] Deloitte Touche Tohmatsu LLC
[6] Visiting Researcher, Institute for Future Initiatives, The University of Tokyo
[7] Toyo University
[8] Deloitte Touche Tohmatsu LLC
[9] CDLE, AI Legal (NEC))
[10] IRCA (International Register of Certified Auditors) "Japan" Members Supporter

* Corresponding author. email: nakano028@toyo.jp

**Summary**

As artificial intelligence (AI) is integrated into various services and systems in society, many companies and organizations have proposed AI principles, policies, and made the related commitments. Conversely, some have proposed the need for independent audits, arguing that the voluntary principles adopted by the developers and providers of AI services and systems insufficiently address risk. This policy recommendation summarizes the issues related to the auditing of AI services and systems and presents a vision for AI auditing that contributes to sound AI governance.

The issues surrounding AI auditing are diverse, and even if the same word is used, different assumptions are made based on different positions and preconditions, making it easy for discussions to be at odds. Therefore, it's important to have common understanding when the parties discuss issues such as audit scope and the timing for putting AI audits into practice.

This policy recommendation addresses the following six issues that are assumed when discussing AI auditing and these are presented in Chapter 2: (1) The need for AI audits, (2) proof proposition, (3) AI audit scope, (4) timing of AI audits, (5) AI audit practitioner requirements, and (6) AI auditing parties and organizations involved.

Chapter 3 examines why AI audits are difficult to conduct from variety of perspective like technical, institutional, and social perspective. Chapter 4 presents the case of recruitment AI to provide specific examples of issues related to AI audits and the reasons why AI audits are difficult to conduct.

In Chapter 5, three recommendations for promoting AI auditing are explained.

Recommendation 1: Development of institutional design for AI audits
Recommendation 2: Training human resources for AI audits
Recommendation 3: Updating AI audits in accordance with technological progress

In this policy recommendation, AI is assumed to be that which recognizes and predicts data with the last chapter outlining how generative AI should be audited.

**About the AI Audit Study Group**
This policy recommendation is the result report of the AI Audit Study Group, a study group of the AI Governance Project, a project of the Technology Governance Research Unit of the Institute for Future Initiatives, The University of Tokyo. The AI Audit Study Group started its research activities in June 2022 and consists of the following members.

Arisa Ema (Associate Professor, Institute for Future Initiatives, The University of Tokyo)
Ryo Sato (Visiting Researcher, Institute for Future Initiatives, The University of Tokyo / Deloitte Touche Tohmatsu LLC)
Tomoharu Hase (Visiting Researcher, Institute for Future Initiatives, The University of Tokyo / Deloitte Touche Tohmatsu LLC)
Masafumi Nakano (Visiting Researcher, Institute for Future Initiatives, The University of Tokyo / The University of Toyo)
Shinji Kamimura (Deloitte Touche Tohmatsu LLC)
Hiromu Kitamura (CDLE, AI Legal (NEC)) / IRCA (International Register of Certified Auditors) "Japan" Members Supporter)

**Process for Developing this Policy Recommendation**

The contents of this policy recommendation were compiled from the discussions of the AI Audit Study Group, consisting of members of the AI Governance Project, the Technology Governance Research Unit of the Institute for Future Initiatives, and external experts. The study group has been meeting online approximately once a month since June 2022 and has also interviewed experts to develop this recommendation. See the Appendix for a list of the study group members and those who cooperated in the interviews.

The policy recommendations will be revised and improved in the future, making effective use of comments from stakeholders and feedback from the webinar scheduled for November 2023 and other related events.

# Table of contents

## 1. Toward AI Audits that Contribute to AI Governance

As artificial intelligence (AI) integrates into various services and systems in society, many companies and organizations have proposed AI principles, policies, or commitments. Conversely, some have proposed the need for independent audits, arguing that voluntary principles by the developers and providers of AI services and systems are not sufficient to address risks. This policy recommendation summarizes the issues related to the auditing of AI services and systems and presents a vision for AI audits that contributes to AI governance. [1]

### 1.1 The Need for Summarizing AI Audit Issues

Discussions on AI audits, audit scope, and timing are so diverse that without a common understanding, the term "AI audit" can lead to misunderstandings in the debate.[2] For example, the assumptions regarding AI technology must be confirmed, such as whether the AI technology to be audited presumes machine learning or, more narrowly, deep learning. Furthermore, contributors' assumptions sometimes differ about the scope of such an audit, e.g., whether it would include the training data sets and external libraries used in addition to the AI technology itself. Other issues, e.g., whether the audit should be conducted at either the development stage or after their release must also be discussed. This policy recommendation discusses machine learning in general, including deep learning, and assumes an AI technology that performs recognition and prediction based on input data. Generative AI, which generates new content and data in accordance with input data and instructions, is introduced and discussed at the end of this paper.

Audits can be categorized as statutory audits based on laws and regulations, such as the audit of financial statements based on the Financial Instruments and Exchange Act, and voluntary audits, which are not legally compelled. Various frameworks govern how voluntary audits are conducted. In Japan, there is currently no statutory audit specifically for AI. However, AI services and systems may be subject to auditing when conducting audits based on other laws and regulations. For example, if an audited entity uses an AI system to prepare financial statement or establish internal controls that rely on the results of judgments made by the AI system, the AI system would also be subject to auditing. Specifically, the design and operating effectiveness of internal controls related to AI, if material, will be verified in the audit of financial statements and audit of internal controls.

The auditing of other AI services and systems, whether internal or external, is voluntary, and the objective, scope, timing, etc. of the audit must be planned by the auditor. Therefore, the understanding of why the audit is necessary, what is to be audited, when, and by whom, must be shared by the parties concerned to realize AI auditing that contributes to AI governance.

### 1.2 Two Trends in AI Auditing

Two major currents have emerged in the auditing of AI services and systems, the first is academic, which focuses on theory, and the other concerns professional practice. This paper

---

[1] The issues surrounding AI and auditing include discussions of (1) auditing AI services and systems, (2) using AI services and systems during audit procedures, and (3) the future of auditing work and the auditing industry (Nakano, 2023). This paper focuses on (1) the auditing of AI systems and not directly (2) on the use of AI services and systems for auditing. Note that a pioneering study on issues (2) and (3) is (Issa et al., 2016).

[2] The definition of AI is not uniformly settled even among experts, but the discussion in this paper is based on this description of AI systems from the OECD's "Expert Group on AI (AIGO)": "Machine-based system that can, for a given set of human-defined objectives, make predictions, recommendations or decisions influencing real or virtual environments. It uses machine and/or human-based inputs to perceive real and/or virtual environments; abstract such perceptions into models (in an automated manner e.g. with ML or manually); and use model inference to formulate options for information or action. AI systems are designed to operate with varying levels of autonomy". (https://doi.org/10.1787/d62f618a-en)

discusses issues arising from these trends.

The first is a discussion that positions auditing of AI services and systems as a means to develop and utilize AI services and systems with trust in the interdisciplinary community that discusses ethics and governance of AI technologies. The latter discusses the difficulties of auditing AI services and systems from the practical perspective of auditors who conduct internal and external audits.

This report results from interviewing experts and other stakeholders and summarizes the issues related to these two trends

## 1.3 Structure of this paper

In Chapter 2, six aspects of AI service and AI system audits are reviewed: 1) the need for AI services and system audits, 2) proof propositions, 3) scope, 4) timing, 5) practitioner requirements, and 6) parties and organizations. Chapter 3 examines the technical, institutional, and social factors that make AI auditing difficult, and Chapter 4 presents a case study in which the contents of Chapters 2 and 3 are applied to an adopted recruitment AI utilized for hiring personnel. Chapter 5 presents future issues and recommendations. Chapter 6 summarizes them. Finally, Chapter 7 introduces issues that should be considered in auditing generative AI, the use of which has been expanding rapidly since 2022.

## 2. Issues Related to AI Audits

In this chapter, after addressing the necessity for AI auditing from the perspectives of AI ethics and AI governance (2.1), the issues of proof propositions (2.2), scope (2.3), timing (2.4), practitioner requirements (2.5), and the parties and organizations involved (2.6) in an audit are reviewed.

## 2.1 The necessity for AI auditing from the perspectives of AI ethics and AI governance

Since the late 2010s, as AI services and systems have expanded, discussions on AI ethics and governance have been debated among various stakeholders in industry, academia, and the private sector.

### (1) Audits required from the AI ethics perspective

AI values and principles have been widely discussed. These include transparency, fairness, and safety (Jobin et al., 2019). Some believe that audits are necessary to assure that AI services and systems comply with these principles. These discussions parallel those on other methods, such as third-party certification and standardization, which form part of a framework enabling users to feel comfortable using AI systems and services. [3] Since risk is assumed in using AI services and systems[4], a provider may have incentives to be audited if their users want such assurance.

In the case of AI systems developed, deployed, and used across organizations or countries in accordance with either domestic or international standards, there is a need for interoperability among frameworks for risk management and auditing in order to translate ethical principles into practice. This "interoperability" between frameworks is also mentioned in Annex 5 of the 2023 G7 Digital and Technology Ministerial Declaration as "interoperability across AI governance frameworks."[5] Unlike mutual recognition or sufficiency certification, which involves the mutual coordination of national processes, promoting interoperability at the "frameworks" level allows the disciplines and responses to AI within each country and organization to coexist and coordinate.[6] This entails discussions and consensus-building regarding various standards related to AI terminology, fundamental concepts, and AI governance and management. Standardization efforts are currently underway through organizations such as ISO/IEC[7], IEEE[8], NIST[9], and CEN-CENELEC[10]. Given that dual management by multiple disciplines may impede innovation and potentially result in societal loss, there is also a need to establish a framework that enables the appropriate operation of these frameworks within each country, region, and application field of AI systems.[11]

---

[3] AI Strategy Council under Japanese Integrated Innovation Promotion Council also states that AI developers and service providers should follow current laws and guidelines to disclose relevant information, to ensure transparency and reliability and that third-party certification and auditing systems should also be referred to https://www8.cao.go.jp/cstp/ai/ronten_honbun.pdf, (p. 10).

[4] For example, the Ministry of Internal Affairs and Communications' "AI Networking Study Council Report 2016" classifies risks into (A) risks of AI not performing its expected functions properly (security risks, risks of transparency and accountability, risks of loss of control, etc.) and (B) risks of AI infringing rights and interests and other legal interests (risks related to the protection of privacy and personal information, risks of being used in crimes, risks related to the rights and interests of consumers, risks related to human dignity and individual autonomy, and risks related to democracy and governance structures). https://www.soumu.go.jp/menu_news/s-news/01iicp01_02000050.html

**(2) Audits required from an AI governance perspective**

Since AI technology is rapidly advancing, existing institutions and governance systems often find it impossible to cope with developments. Therefore, a governance approach has been proposed, which allows institutions and systems to be agilely updated. Internal and external audits are proposed to assess the credibility of agile governance systems.[12]

A characteristic of AI services and systems is that they are regularly updated by retraining after implementation. In this case, monitoring process to confirm whether they work properly and maintenance process in the case of any trouble are required. AI services that are not properly managed may become inaccurate and amplify the social risks of misjudgments. Thus, in addition to effective governance during their preparation of services and systems, such audits are also needed during the operation and maintenance phases of AI services and systems.

---

[5] Results of G7 Digital and Tech Ministers' Meeting in Takasaki, Gunma, Annex 5: "G7 Action Plan for promoting global interoperability between tools for trustworthy AI",
https://www.meti.go.jp/press/2023/04/20230430001/20230430001-ANNEX5.pdf

[6] For example, OECD compares standards of ISO, IEEE, and NIST, EU AI Act and Rule of law impact assessment from Council of Europe (HUDERIA) as interoperability framework. OECD.AI work promoting interoperability of AI risk management frameworks, IGF Policy Network on AI meeting #4, 18 July 2023, https://www.intgovforum.org/en/filedepot_download/282/25999
OECD also classifies AI system and AI system lifecycle and provide framework for comparison, Advancing accountability in AI, https://doi.org/10.1787/2448f04b-en.
In Japan, the Ministry of Economy, Trade and Industry (hereinafter called "METI") published "Governance Guidelines for Implementation of AI Principles",
https://www.meti.go.jp/press/2021/01/20220125001/20220124003.html

[7] ISO/IEC JTC 1/SC42 has so far already prepared and published 20 international standards, including AI concepts and terminology (ISO/IEC 22989) and Governance implications of the use of AI by organizations (ISO/IEC 38507). 31 international standards are also under discussion.

[8] IEEE 7000 series and others discuss standards for practical issues in AI.

[9] NIST discusses the integrated risk-based framework for AI that is interoperable with ISO/IEC management standards/concepts and OECD Recommendation of the Council on Artificial Intelligence and also organizes their relationships.

[10] CEN/CENELEC discusses standards for AI in Europe.

[11] For a discussion on interoperability among these frameworks, see the policy recommendation published by Institute for Future Initiatives, The University of Tokyo "Towards Responsible AI Deployment Policy Recommendations for the Hiroshima AI Process" (https://ifi.u-tokyo.ac.jp/en/wp-content/uploads/2023/09/policy_recommendation_tg_20230915e.pdf). Institute for Future Initiatives, The University of Tokyo also provides a framework for AI governance called Risk Chain Model, which can be used as a preliminary work for AI services and system audits. Actual case models are available in their website, https://ifi.u-tokyo.ac.jp/en/projects/ai-service-and-risk-coordination/. Policy recommendations "RCModel, a Risk Chain Model for Risk Reduction in AI Services" is also published, https://ifi.u-tokyo.ac.jp/en/wp-content/uploads/2020/07/policy_recommendation_tg_20200706.pdf.

[12] It is also discussed in METI "Agile Governance Update -How Governments, Businesses and Civil Society Can Create a Better World By Reimagining Governance-" (2022) and other publications.
https://www.meti.go.jp/press/2022/08/20220808001/20220808001.html

Furthermore, a balance is required between the need and the costs for each service undergoing an AI audit. One strategy is the risk-based approach, which determines that the level of audit relates to the risk associated with the target AI service and that the audit procedures and costs are graded accordingly. For example, the European Artificial intelligence act (hereinafter called "EU AI act") adopts a risk-based approach and states that independent auditor reporting must be included for high-risk AI systems. [13]

## 2.2 AI Audit Proof Propositions

In this paper, we use the term "proof propositions" to describe what should be considered audit topics when conducting an audit.[14] The proof propositions for AI audits include items that appear in the AI Principles, in various guidelines, and items that are considered as proof propositions in system audits other than AI systems.[15] Among them, those that are unique to audits for AI services and systems are listed in Table 1.

**Table 1 Examples of AI Audit Proof Propositions**

| Proof propositions | Explanation |
|---|---|
| Fairness | Is there any inappropriate bias in the output results of the AI system, etc.? <br> It is required to have a common understanding of the definition of fairness beforehand. |
| Transparency | Can the output results of the AI system be reproduced, and can the training data and feature values be explained, etc.? |
| Safety | Is there any possibility that the AI system may harm the user? If a problem occurs, does the system properly transition to a halted state etc.? <br> Hardware in which AI systems are embedded should also be considered. |
| Security | Can attacks on training data be prevented or detected? Can production input data that intentionally induce inappropriate output be prevented etc.? |
| Privacy | Can individuals refuse to attribute data they do not wish to share? <br> Can erroneous personal assessments be corrected in a timely and appropriate manner etc.? |

When conducting audits for these proof propositions, some of the standards, criteria, and frameworks used in existing system audits can be applied. For example, security and privacy can be covered to a certain extent using the trust services criteria adopted in existing SOC2 reports.[16] Conversely, taking into account the technical complexity of AI, the wide range of parties involved, and considerations unique to AI (see 3.1), it is difficult to address all proof propositions in an AI audit using only existing criteria.

---

[13] Council of the European Union, Proposal for a Regulation of the European Parliament and of the Council laying down harmonized rules on artificial intelligence (Artificial Intelligence Act) and amending certain Union legislative acts - General approach (2021), https://eur-lex.europa.eu/legal-content/EN/TXT/?uri=celex%3A52021PC0206

[14] The term "subject of the audit" would cause confusion with the individual propositions to be proved as an audit listed in 2.3, so the term "proof proposition" of the audit is used here. Toba (2009) separates the object of audit from the subject of audit by the concept of "subject matter" of the audit and makes a sharp distinction from the audit object that appears uniquely in various forms in the audit process.

[15] The Ministry of Internal Affairs and Communications' "AI Network Society Promoting Council Report 2022 " (https://www.soumu.go.jp//000826564.pdf) identifies 22 values to be respected based on the principles, policies, and guidelines of each country.

**2.3 AI Audit Scope**
The AI systems and services' audit scope can be classified into two categories: auditing individual AI services and systems in operation themselves (2.3.1) and auditing the internal controls implemented in organizations that provide AI services and systems (2.3.2). The former case audits the services, systems, models, and the programs themselves, while the latter case audits governance, management processes, rules and policies, human tasks, chain of thought, etc. within the organization. Since the perspectives, methods, and procedures differ greatly between the two types of audits, if an audit is conducted without specifying and agreeing in advance on which parts are to be examined, the audit results will not meet the expectations of all parties involved. Since these categories are set for the sake of convenience, auditors may combine them in their actual audit process.

**2.3.1 Auditing AI services and systems**
To audit an AI services or systems, AI system and services in their audit scope must be identified. Although trained models are main component of AI system, components other than the trained models will also be subject to auditing in AI systems and relevant services audit. Figure 1 shows a diagram of AI services and systems broken down by the components considered to be subject to auditing.[17] The following explains the components, starting from the outer frame.

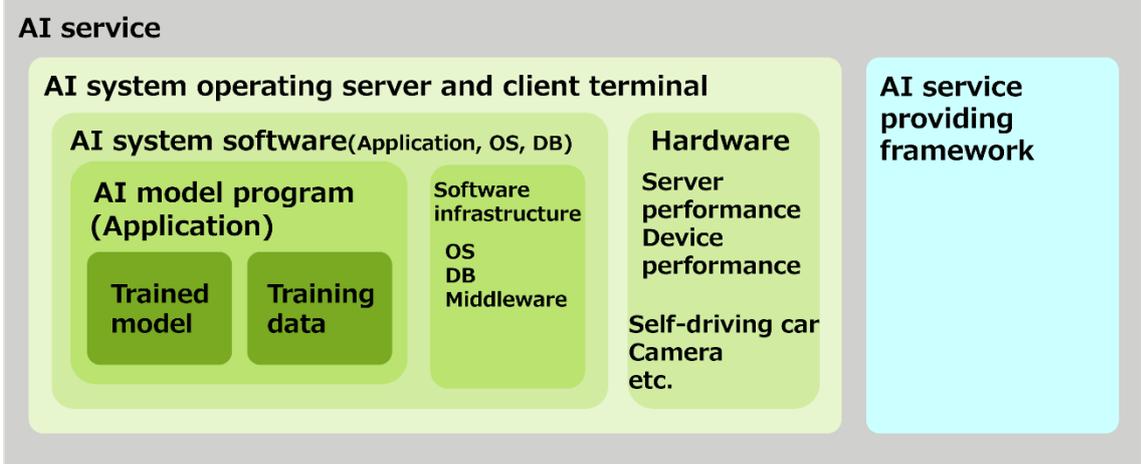

**Figure 1 Components of AI services and systems**

---

[16] A SOC 2 report is a report that expresses the results of the auditor's procedures and opinion on internal controls related to security and other topics described by a service organization following the Trust Services Criteria established by The American Institute of Certified Public Accountants. Other standards and frameworks such as The Japanese Institute of Certified Public Accountants (hereinafter called "JICPA") "Japanese Standard on Auditing 315 Identifying And Assessing The Risks Of Material Misstatement" (https://jicpa.or.jp/specialized_field/2-24-315-2-20230810.pdf) and METI "System Auditing Standards"(https://www.meti.go.jp/policy/netsecurity/sys-kansa/sys-kansa-2023r.pdf) are also considered to be applicable.

[17] Prepared with reference to JICPA " Japanese Standard on Auditing 315 Identifying And Assessing The Risks Of Material Misstatement" (https://jicpa.or.jp/specialized_field/2-24-315-2-20230810.pdf), IPA "Common Framework 2013", Jakob Mökander et al. (2023)

**(1) Components of AI Services**

An AI service consists of an AI system with a trained model and a framework within which the organization provides the service. When examining the organization's service framework, e.g., whether the organization discloses that it uses AI, and that information used by the service is handled appropriately, are examples that may be subject to audit depending on the audit objective. [18]

On the other hand, entire AI systems could be subject to audit, including hardware such as the operating enclosures and terminals in which the trained models are stored[19]. Additionally, AI systems often comprise (i) a server on which the trained models are stored and (ii) clients that give instructions to the server or display the results of the AI's decisions, and both server and client terminals can be subject to audit depending on the audit objective.

**(2) Components of AI system operating enclosures and terminals**

In addition to software including the AI model's programs, the performance of the AI system's operating enclosures and terminals, as well as hardware such as the AI system's operating servers, client terminals, and devices necessary for providing services may also be subject to audit, depending on audit objective. For example, when auditing an AI biometric authentication system's performance, its overall performance could be affected by the performance of devices such as cameras, so these devices should also be considered as audit scope. In the case of edge computing, the processing performance of the server and how its location affects the processing speed (latency), and in the case of using cloud services for software infrastructure, each customized service usage and configuration setting and regional selection may also be subject to auditing regarding security and availability.

**(3) AI System Software Components**

AI model programs (applications) that contain trained models are the main audit scope, but other software infrastructure such as operating systems (OSs), databases (DBs), and middleware can also be audited. These components can be subject to auditing similarly to current system audits.

**(4) Components of the AI Model Program**

The auditing of AI model programs includes research on algorithm auditing (Bandy, 2021). Algorithm auditing can be defined as the study and practice of evaluating, mitigating, and assuring the legality, ethics, and safety of algorithms (Koshiyama et al., 2022).

Importantly, the training data used to make the trained model could also be included in the audit. The data collection process, the sufficiency of data regarding outliers and biases, and the method of setting correct labels are also potential audit items (Batarseh et al., 2021).

**2.3.2 Audits of internal controls in organizations that provide AI services and systems**

The internal controls implemented in an organization that provides AI services and systems, are audited by focusing on governance, management processes, rules and policies, human tasks, and the chain of thought within the organization, and whether these are properly designed and operated. This section summarizes the contents, methods, scope, and targets of internal control audits, assuming that they are conducted within the same framework as existing audit of internal controls.

---

[18] The standard for IT service management is ISO/IEC 20000 IT Service Management System.
[19] In METI "System Management Standards", Chapter II.2.7, hardware is also listed as a component of information systems.
https://www.meti.go.jp/policy/netsecurity/sys-kansa/sys-kanri-2023.pdf

It is assumed that internal controls are generally established and audited in accordance with the existing frameworks such as COSO[20] and COBIT[21]. However, due to the factors listed in Chapter 3 that make AI auditing difficult, certain issues cannot be addressed within the existing frameworks only.

**(1) Contents of Audits**

Examples of existing audit for organizational internal control include an audit of internal controls by external auditors, internal audits by corporate internal auditors, and management system audits based on Guidelines for auditing management systems (ISO 19011)[22]. In all of these, the scope of the audit is whether the organization appropriately designs rules and whether these rules are operated effectively by each personnel performing tasks following them. The audit scope includes processes, human tasks, and chains of thought. When AI services and AI systems are introduced within an existing framework of audits, audits of the organization and internal controls related to the AI services or systems are considered necessary as part of these audit frameworks. In addition, AI governance auditing may form an extension of IT governance and IT management. An AI compliance audit may be conducted to examine whether the AI system and service are appropriately managing personal information and privacy.

**(2) Audit Methodology**

The internal controls of AI system and service organizations are also expected to be in line with internal control frameworks such as COSO and COBIT described above. When auditing internal controls, the typical method is to confirm rules are established to reduce the risk which each organization face and to satisfy the audit's proof proposition and these rules are properly operated by organization personnel through interviews and inspection of the relevant audit evidence. [23]

**(3) Audit Scope and Target**

In general, the scope of the control activities subject to an audit of internal controls ranges from entity level control activities covering the entire company to control activities performed on individual services and systems. This is assumed to be the same for AI services and systems.

If the entity level controls, in other words, the entire organization level control activities which should be managed at the management level are to be examined, the establishment of guidelines for AI utilization in the company, the education and training system for AI use, etc., may be considered as audit scope. Conversely, if control activities at the level of individual AI service or system personnel are to be considered, examples of control activities to be audited include the preparation of test plans, approval of test results, and related tasks and procedures prior to publicly releasing the AI services or systems.[24] These control activities are considered to include both those similar to control activities for non-AI services and systems and those unique to AI services and systems.

---

[20] An internal control framework proposed by the Committee of Sponsoring Organizations of the Treadway Commission. It consists of the following elements: Control Environment, Risk Assessment, Control Activities, Information and Communication, and Monitoring Activities.
[21] ISACA, ITGI's framework for entity-wide information and technology governance and management for business entities, is derived from Control Objectives for Information Technologies.
[22] For more detail on ISO19011, refer to https://www.iso.org/standard/70017.html
[23] More detailed methods are introduced in Financial Service Agency "Standards for Management Assessment and Audit of Internal Control Over Financial Reporting".
[24] System-related internal controls over general systems are introduced in JICPA "Japanese Standard on Auditing 315 Identifying And Assessing The Risks Of Material Misstatement" (https://jicpa.or.jp/specialized_field/2-24-315-2-20230810.pdf).

## 2.4 Timing of AI Audits

The timing of AI audits varies by audit scope and type (internal or external audit). In this section, after classifying the lifecycle of AI services and systems, audit timing is organized by audit scope and type.

### 2.4.1 Classification of AI Life Cycle

Referencing the OECD definition of the AI lifecycle[25] and general systems' development lifecycles[26] and to consider the discussions concerning each audit scope and type, this paper classifies the AI lifecycle into four phases: (1) new development, (2) functional change or additional development, (3) operation, and (4) disposal (Figure 2).

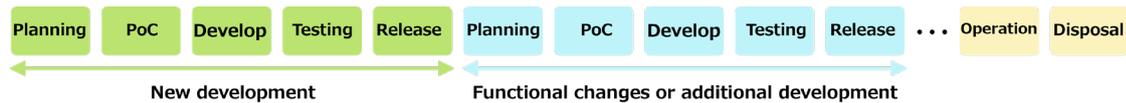

**Figure 2: AI Life Cycle**

**(1) New development**

Before releasing a new AI service or system, the concept is defined in the planning phase, and after a proof of concept (PoC), the decision to proceed or not is made. After that, the system proceeds to the development and testing phases. Once the test results confirm that the quality specified in the specifications has been met, the AI service or system can be released.

**(2) Functional changes and additional development**

When the functionality of an AI service or system already in operation must be changed or to implement additional functions, it is released through the necessary processes from the planning phase to the testing phase again depending on the scale and requirements. Since the design and development of AI services and systems assume that the accuracy of models deteriorates due to data fluctuations and that the algorithms used to prepare trained models are subject to change, functional changes and additional development are frequent. [27]

**(3) Operation**

For AI services and systems in operation, batch jobs, error monitoring, and responses to failures must be executed. This operational phase can be variously defined. Not only monitoring and maintenance process, but also the functional changes and additional development process described in (2) can be considered aspects of the operational phase.

**(4) Disposal**

After the AI services and systems go through the convergence processes and procedures for termination, the final phase is appropriately disposing of the retained data and programs and disseminating the required information to users.

### 2.4.2 Classification of audit timing by scope

As in 2.3, AI audits can be categorized as either focus on AI services and systems or the organizations that provide AI services and systems and their internal controls.

---

[25] As the lifecycle of an AI service or system, the OECD categorizes it into four phases: (1) Design, data and models; (2) Verification and validation; (3) Deployment; and (4) Operation & Monitoring. OECD. "Scoping the OECD AI principles: Deliberations of the Expert Group on Artificial Intelligence at the OECD (AIGO)", OECD Digital Economy Papers, No. 291, OECD Publishing, Paris, https://doi.org/10.1787/d62f618a-en.
[26] IPA "Common Frame 2013" introduces a more detailed process, from the stakeholder requirement definition process to the software disposal process.
[27] Ichiro, Akimoto et al. "AI Business Compendium" (President, Inc.)

**(1) Timing of audits of AI services and systems**

Current system audits, especially external audits, are often conducted for the scope system in production environment after a new development is released. However, when auditing an AI service or system based on the inductive method to decipher patterns and trends from a large amount of data, the validity of the AI service itself and also the validity of decisions on whether or not development is necessary. Therefore, retrospective audits may be conducted for pre-release phase, including the testing, development, PoC, and planning phases, if necessary.

It should also be noted that, especially in the case of AI services or systems that continuously being trained, the accuracy may differ between the AI output results at the timing of the audit procedure and AI output results when such results are used. For example, even if the same input data is used, the output results may change both before and after additional learning, resulting in different performance indices, such as accuracy rates or precision rates.

**(2) Timing of audits of internal controls in organizations that provide AI systems and services**

For an organization that provides AI services and systems, the internal controls spanning the entire lifecycle from planning to disposal are assumed to be audited.

When it comes to timing of AI audit, in the case of external audits, the first year of operation is considered when the scope service is released. However, in the case of continuous audits or when multiple services are in scope, evaluations are expected to be conducted throughout the year. In practice, it is reasonable to align the period with other audits, and annual intervals (usually one year) are generally considered to be the normal period. However, because AI technology evolves rapidly, and depending on the degree of risk, audits may be conducted at an appropriate frequency not limited to this timeframe.

### 2.4.3 Classification of audit timing by type

There are two types of entities that conduct audits: internal and external based on who performs the audit, and they differ in their timing.

**(1) Internal Audit**

The audit objectives and audit objectives can be set arbitrarily for internal audits. Therefore, all phases of the lifecycle can be audited, or specific phases can be audited in depth. For example, from the planning phase, an internal auditor may work with the relevant department and conduct an audit to improve the quality and governance of an AI system and service and design a system to facilitate future audits.

**(2) External Audit**

Currently, external audits are conducted mainly after the development phase.[28] However, in providing AI services, the appropriateness of the service itself and the appropriateness of decisions on whether or not development is necessary are also important issues. Therefore, external audits that include the planning and PoC phases may be considered necessary.

### 2.5 AI Audit Practitioner Requirements

In this section, the AI audit practitioner requirements are presented from the perspectives of (1) expertise requirements, (2) independence requirements, (3) organizational requirements, and (4) legal responsibilities of the auditor.[29] Of these, the expertise requirement (1) and independence requirement (2) are common to both internal and external audits, but external audits additionally include the organizational requirements (3) and the auditor's legal responsibility (4).

---

[28] However, this does not necessarily apply to other than system audits, such as quality management system (QMS) audits.

### 2.5.1 Expertise Requirements

Conducting an AI audit requires diverse expertise with a wide range of skills and experience. An understanding of audit theory, industry knowledge related to the audited company and the services it provides, and knowledge and experience in IT areas not limited to AI, deep AI-specific technical knowledge, and ethical, cultural, legal, and regulatory expertise are all necessary. Thus, the requirements for those conducting AI audits are very high, and an individual is unlikely to possess all the skills and experience required for AI auditing. Realistically, therefore, audits will be conducted by teams of practitioners. In addition, when conducting a full-scope audit covering areas other than the AI system and service, depending on the design of the institutional design for audit, the AI audit practitioner may be considered an expert, and the audit may be conducted under the "Using the Work of an Auditor's Expert" scheme in the areas related to the AI system and service.[30] However, the size of the audit team would be determined only after considering the cost can be used in an AI audit.

### 2.5.2 Independence Requirements

As with traditional audits, AI audits must be conducted by independent practitioners and organizations that have no conflict of interest with the audited company or department. Independence requirements are defined for both external and internal audits. [31]

### 2.5.3 Organizational Requirements

To ensure reliability in the AI audit results, the organization conducting the audit must meet certain quality and independence standards. By implementing an organizational structure to ensure audit quality and independence of AI auditors and monitoring the effectiveness, AI users can use the audit results with reliance.

For an organization that meets the above organizational requirements to exist, the certification and accreditation system for the organization, as well as the organization and its role in conducting monitoring, must be considered. Notably, the audit results conducted by an organization that does not meet these standards may not credibly reflect the actual situation.

### 2.5.4 Legal responsibility of the auditor

Although an AI audit practitioner has conducted an audit with due care,[32] the audit results may not be correct. In such cases, the scope of the AI audit practitioner's legal liability must be considered[33]. For example, if an audit expressed an unqualified opinion on the safety of an AI system installed in self-driving car, but an accident was caused due to the AI system, the auditor's scope of responsibility must necessarily be considered.

---

[29] The requirements in Business Accounting Council, Financial Services Agency "Auditing Standards" (https://www.fsa.go.jp/singi/singi_kigyou/kijun/20201106_kansa.pdf) and The institute of international Auditors Japan "Japanese Internal Auditing Standards" (https://www.iiajapan.com/leg/pdf/guide/20140601_2.pdf) are summarized.

[30] The use of experts in external audits is governed by JICPA "Japanese Standard on Auditing 620 Using the Work of an Auditor's Expert " (https://jicpa.or.jp/specialized_field/2-24-620-2-20230810.pdf).

[31] In external audits, JICPA "JICPA Code of Ethics for Professional Accountants" (https://jicpa.or.jp/specialized_field/files/2-22-0-2-20221031.pdf) requires certified public accountants to be both independence of mind and independence in appearance (Section 120.15.A1). Also, in internal audits, in terms of independence and objectivity, the internal audit activity must be independent, and internal auditors must be objective in performing their work (IPPF 1100, IIA (2017d)).

[32] Under JICPA "JICPA Code of Ethics for Professional Accountants" (https://jicpa.or.jp/specialized_field/files/2-22-0-2-20221031.pdf), "Professional competence and due care requires the professional accountant to have professional knowledge and skill at the level required to ensure the provision of competent professional service, and to act diligently in accordance with applicable standards, laws and regulations…" (Section 120.16 A2(3)).

If the responsibility of AI audit practitioners is too burdensome, very few would be willing to conduct the audits. Therefore, the need for exemption requirement or insurance system to protect AI audit practitioners should be considered.

## 2.6 AI auditing Parties and Organizations Involved

Auditors and audited companies are not the only parties involved in auditing AI services and systems. Figure 3 summarizes the parties and organizations involved in AI auditing as currently envisioned. The following sections describe (1) AI service provider organizations, (2) AI service users and user organizations, (3) external auditors, (4) standardization bodies/certification and accreditation bodies, (5) public institutions, (6) private organizations, and (7) other related parties.

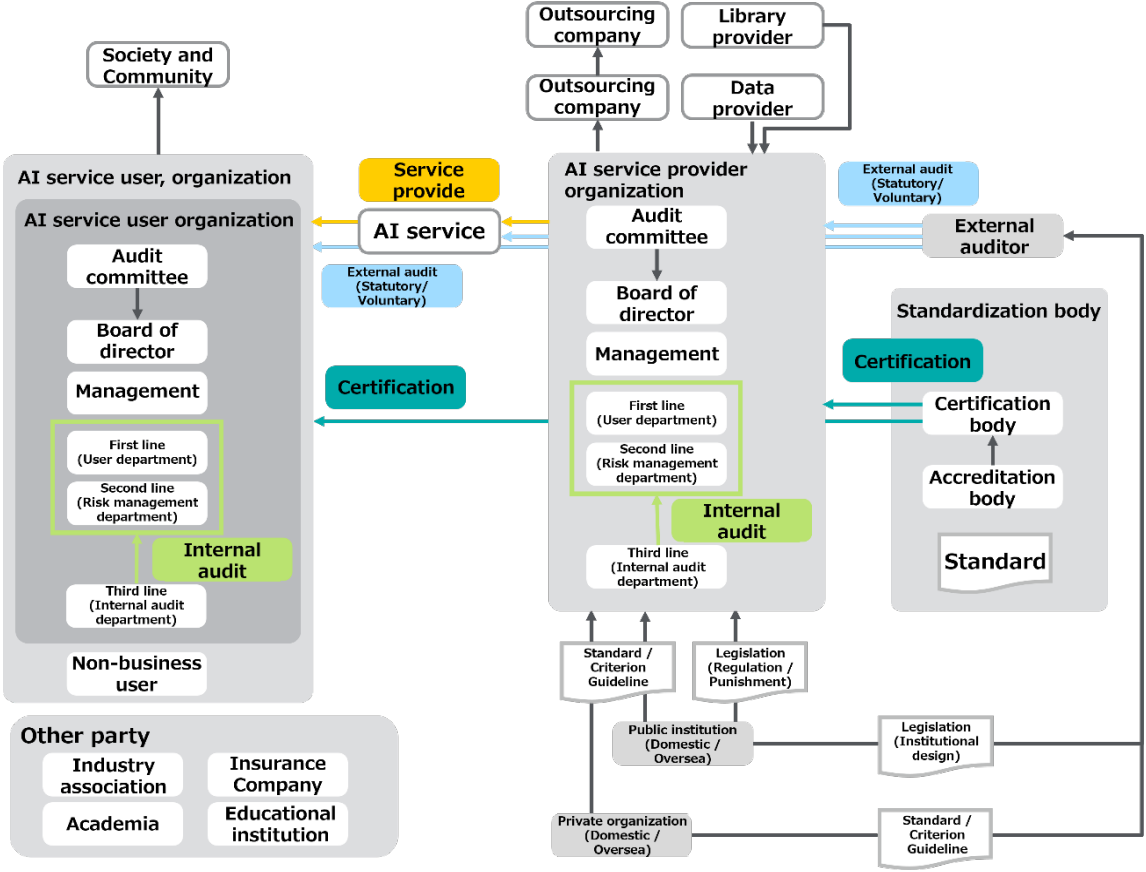

**Figure 3: Parties and organizations involved in AI audits**

### 2.6.1 AI Service Provider Organizations

AI services and systems are expected to be provided by organizations individually or across organizations.

---

[33] In JICPA "JICPA Laws and Regulations Committee Research Report No. 1, Legal Liability of Certified Public Accountants, "(https://jicpa.or.jp/specialized_field/files/2-15-1-2-20160801.pdf) the liability of certified public accountants is explained under four categories: civil, administrative, criminal, and other.

**(1) Governance within the organization**

Organizations that provide AI services or systems are subject to audits of such services, or systems, or of the internal controls in organization, as categorized in 2.3 AI Audit Objectives.

The Institute of Internal Auditors has proposed a three-line model for proper governance, including organizational risk management.[34] The three lines for an organization providing AI services and systems are: 1) the department that develops and operates the AI system and service, and 2) the risk management department that checks and balances the development and operation of the AI services and systems. Management and the executive department, in their roles as the first and second lines, aim to enhance and improve their governance and management. The third line 3) is the internal audit department that independently verifies and audits the management and executive departments.[35] Whether the three-line model effectively works is expected to one of the audit focuses by audit committee (Using external auditor, if needed).

**(2) Governance across organizations**

As the complexity and sophistication of supply chains increase, it is often impractical to build AI services and systems with only in-house resources, thus building them could involve external providers. Whether and to what extent to include outsourcing companies, external libraries, and data providers that support system construction as entities to be audited must be considered. In addition, the relationships among AI services and systems' providers and outsourcing companies are diverse. Depending on the relationships, audit process needs to be considered from different perspective. The following illustrates some possible cases:[36] .

A) Companies that, using both in-house resources and outsourcing companies, have established an AI system and service for external users.
B) Companies that outsourced the development of an AI system and service, and use it only internally
C) Software vendor companies that build AI systems for clients and sell packages

**2.6.2 AI service users, user organizations**

As well as the service providers, AI service users and user organizations may also be audited in their position as users. In this case, the main audit focus would be how they use AI services and utilize the results of AI decisions within their organizations. Therefore, the first line of the three-line model would be departments that use the AI services.

When an AI service is built and used by the same company, the audit should be conducted from the perspectives of the service provider (2.6.1) and the AI service user (2.6.2).

**2.6.3 External Auditors**

An independent third party who conducts an external audit of an audited entity shall perform verification from a completely independent position according to the laws and regulations and shall disclose the audit results as an audit opinion. Issues to be considered in external auditor's case include the expertise requirements, independence requirements, organizational requirements, and legal responsibility of such an auditor (see 2.5).

---

[34] The three-line model was published by IIA as a model for the risk management and governance of organizations in 2020 (IIA (2020)).
[35] In addition to internal audits, audits by audit supervisory board members may also be considered from the standpoint of three-party audits, but we omit a detailed discussion of these in this report.
[36] Referring to METI "Contract Guidelines for the Use of AI and Data,"
(https://www.meti.go.jp/policy/mono_info_service/connected_industries/sharing_and_utilization/20180615001-1.pdf) the report assumes and presents cases that are considered representative.

### 2.6.4 Standardization Bodies/Certification and Accreditation Bodies

The standards organizations that establish criteria and standards for internal and external audits include the International Organization for Standardization (ISO), the International Electrotechnical Commission (IEC), the Institute of Electrical and Electronics Engineers (IEEE), the Japanese Industrial Standards Committee, and others. The criteria and standards established by these organizations are used or referenced in audits. The third parties that audit the conformity to ISO standards are called certification bodies, and in Japan, an accreditation body (JAB) examines the certification bodies.

### 2.6.5 Public institutions

A means of developing institution design for AI audits is through the legislation passed by legislatures. Therefore, domestic and foreign public institutions, including legislative bodies and representatives, should be considered as stakeholders in AI audits.

Conversely, the specific procedures for conducting AI audits, judging audit results, and the rules and procedures applicable to AI service providers are expected to comply with publicly available standards of practice. Sometimes public institutions prepare these rules, guidelines, and standards. For example, METI "System Audit Standards"[37] and "System Management Standards,"[38] which were revised in April 2023, are Japanese standards for system audits.

### 2.6.6 Private organizations

Many of the standards that AI service provider organizations and AI audit practitioners must comply with are issued by private organizations. For example, in an audit of financial statements, accounting standards which the entity being audited must comply are developed and published by the Accounting Standards Board of Japan (ASBJ), a private organization. Therefore, in AI auditing, the domestic and foreign private organizations that issue such standards should also be considered parties to the process.

### 2.6.7 Other Parties

Other stakeholders include academia and educational institutions in terms of continuous innovation and user education, industry associations in terms of regulation and coordination specific to the individual industries providing AI services, and insurance companies in terms of defining insurance systems as risk controls from a different perspective than auditing. These parties are directly or indirectly related to the audit of AI services, and therefore, should be considered as a related party of the AI audit.

## 3. Factors that make AI Audits Difficult

Conceptual, technical, institutional, and social factors have been identified that make auditing AI services and systems difficult (Mökander et al. (2023)). This paper focuses on five of these factors: (1) the complexity of AI technology, (2) the underdevelopment of institutional design for AI audits, (3) the difficulties in setting performance standards for conducting AI audits, (4) the complexity arising from the scope of audited entities, and (5) the imbalance between demand and supply for AI audits.

### 3.1 Complexity of AI Technology

In current system audits, systems are designed, programmed, and implemented according to the correct answers defined by business requirements. Their accuracy and other aspects are evaluated based on whether the system outputs the values uniquely produced by the input data, the established logic, and by additional verification methods. Unlike current IT systems, however, AI systems are not designed to output pre-defined unique values.

---

[37] METI. "System Audit Standards", https://www.meti.go.jp/policy/netsecurity/sys-kansa/sys-kansa-2023r.pdf
[38] METI. "System Management Standards", https://www.meti.go.jp/policy/netsecurity/sys-kansa/sys-kanri-2023.pdf

In the development of AI systems, the logic of decision-making is often black boxed, and it is difficult to verify the logic. Even if the source code of the program and parameters such as feature values are disclosed and the logic can be reproduced by mathematical formulas or conditional judgment statements, the formulas and parameters are expected to be computationally complex, hampering a full assessment of validity.

Furthermore, depending on how the AI system is constructed, the model may be trained by including random elements. In such cases, even if an identical training method is used, different trained models can be built. Therefore, audit decisions must be made to include such random elements.

In addition, as mentioned in the timing of AI audits (2.4), some AI services and systems conduct continuous learning. In this case, an evaluation at the time of the audit may differ from an evaluation when the audit results are used, making it difficult to conduct the audit. It is also difficult to determine whether the output has changed due to continuous learning or a bug, which increases the difficulty of AI auditing regarding reproducibility.

Finally, the complexity of human-machine interactions also increases these difficulties. AI services and systems can be designed for human decision-making support, or for making decisions without human intervention. In domains such as medicine, where AI systems are used as diagnostic aids and the final decision is made by the physician, the role of auditing depends on the assumptions and context in which the technology is used.

### 3.2 Underdevelopment of institutional design for AI audits

By September 2023, AI-specific general standards had not yet been established for audit practitioner requirements and quality management systems. Additionally, no standardized criteria or standards of practice (performance standards) had yet been established for audit procedures, leaving auditors to design their own procedures and judge the audit results (Akoshima & Fukuda 2020a, 2020b, 2020c).

Furthermore, there are no unified rules or firm agreements concerning with which domestic and international rules, management, auditors, certification bodies, and other parties must comply. Consequently, these parties do not fully understand their compliance obligations. This immaturity in institutional design for AI audits is one of the factors that vexes the implementation of AI audits.

### 3.3 Difficulties in Setting Performance Standards for AI Audits

Even when auditors formulate their own procedures, setting performance standards for AI audits is difficult. For example, various proof propositions (2.2) are envisioned for AI audits. Among them, some, e.g., fairness, are difficult to define precisely and difficult to verify. To illustrate, there are two major technical fairness indicators[39], but they are not intended to be achieved simultaneously.

Furthermore, trade-off relationships between different values must be considered. In general, there is a trade-off between ensuring the accuracy of AI judgments and accountability. Trade-offs can also be assumed between the audit's proof propositions. An example is the difficulty of conducting an audit that encompasses multiple perspectives such as the perspectives of security and accuracy and that of efficiency, because the audit issues and verification items differ.

---

[39] One is "individual fairness," which refers to the state in which one individual is treated like any other individual, regardless of group attributes, and the other is "group fairness," which refers to fairness among sensitive groups such as men and women within a group. The main criteria for this group fairness are statistical parity (Dwork, 2012) and equalized odds (Hardt, 2016).

In addition, when conducting an AI audit, the data governance issues related to training data must also be examined. However, the potential biases in such data and the difficulty of setting sufficiency indicators would make it difficult to conduct AI audit from the viewpoint of performance standards for AI audit.

### 3.4 Complexity arising from the scope of audited entities

As introduced in the section on AI auditing parties and organizations involved (2.6), the AI system and service have many stakeholders, making it difficult to define the scope of who should be included in the audit.

### 3.4.1 Development and Operation of AI Systems Across Organizations

One of the characteristics of AI systems and services is that they are developed and used across countries and organizations. It will become difficult to conduct AI audits if they necessitate the identification of multiple organizations, such as outsourcing companies, as audit scope and then conduct audits across organizations, rather than within a single organization.

For example, if external vendors contribute to a system's development and operation, the extent of the vendors' inclusion in an audit scope must be considered. If the vendor is the main entity implementing internal control during development and operation, and the principal company is not involved, the vendor should be treated as a service organization, and the internal control implemented by the vendor should be included in the audit scope.

Conversely, when auditing a service organization, the issue of determining the audit scope arises. In addition, there is a case that a vendor subcontracts to another vendor. Practically speaking, because of the parties' contractual obligations, auditing multiple outsourcing companies will be difficult. Even if audits can be conducted, there may be uncertainty about who carries the costs of auditing the outsourcing companies.

### 3.4.2 Use of External Libraries

In many cases, AI system developers use publicly available functions from external libraries. In such cases, the correct operation of an AI system could rely on the functions and processes provided by external libraries. If the external library functions and their providers are included, the audit scope becomes even broader, without the assumption that the functions provided by these libraries are functioning properly.

### 3.4.3 Training Data Governance

Many stakeholders, such as data collectors, public data providers, and data labelers for training are involved in the management of training data. Because the validity of training data is essential in considering an AI audit, the audit scope becomes even broader when it includes such stakeholders and their governance.

### 3.5 Imbalance between demand and supply for AI audits

As summarized in (1.1), the need for auditing AI services and systems is increasing as they become more prevalent. However, conducting AI audits under the appropriate scope and AI audit practitioner requirements is difficult due to diverse expectations concerning the assured outputs of AI audits, the lack of human resources, and the absence of incentives among other factors.

### 3.5.1 Mismatched Expectations regarding the assurances offered by AI Audits

As summarized in 2.3, the scope of AI audits can be divided into two categories: The AI

systems and services themselves, and the internal controls implemented by the provider organizations. In some cases, it may be difficult to audit AI services or systems themselves due to the technological, institutional, or social factors described in Chapter 3. In such cases, the AI audits will only provide assurances that the provider organizations' internal controls are effective. In addition to such audit limited scope, there are various restrictions and limitations on the content of such assurance.[40] In contrast, if society expects assurances for AI systems and services, e.g., "AI services must operate 100% correctly and be safe," and if the characteristics of AI services and systems are not considered with regard to such expectations, a gap will develop between the requirements of AI audits and their implementation.

### 3.5.2 Lack of human resources to conduct AI audits

As summarized in the AI Audit Practitioner requirements (2.5), the conducting AI audits requires expertise in auditing and a wide range of knowledge covering specific knowledge of audit and technical knowledge of AI, as well as legal and ethical content. At present, however, it is assumed that there is a shortage of personnel capable of conducting such audits.

### 3.5.3 Mismatch between supply and demand for AI auditors

If the audit results are incorrect, the audit practitioner could be held legally liable. However, as introduced in Chapter 3, technical, institutional, and social factors make it difficult to conduct audits, and the auditor's responsibility and audit fees may become imbalanced. Therefore, even if the demand for AI audits increases, there is little incentive for external auditor to undertake such work.

### 3.5.4 Lack of incentives for audited companies to undergo AI audits

Not only is there little incentive for auditors to conduct audits, but there is still little demand for AI audits from audited companies. By September 2023, as far as this study group could determine, there were not yet legally binding regulations or penalties in Japan that would make AI audits mandatory. There were neither regulations nor penalties could be observed at the national levels except for state levels or ordinances.[41] In addition, as summarized in Chapter 3, since they cover a wide range of topics, AI audits are difficult to conduct and high-quality audits involve significant costs, further disincentivizing companies from undergoing audits.

### 4. Assumed AI audit case study: Recruitment AI

The specific example is given of a recruitment AI service to introduce the issues surrounding AI audits described in Chapter 2 and the factors that make AI audits difficult described in Chapter 3. The recruitment AI case study here assumes a service that makes hiring decisions based on job applications and interview recordings. Although there may be some issues that overlap with those to be considered in a non-AI system audit, we will focus on issues that are characteristic of AI audits when applied to AI services and systems.

### 4.1 The Need for AI Audits

For a company that uses recruitment AI services, not only the preciseness of the AI's judgment, but also the fairness and other aspects are considered and assured in their audit, will reassure adopters and enhance the company's reputation. A recruitment AI is considered a high-risk AI in the EU AI act and service providers are required to undergo a conformity assessment in advance. If this act is enforced, audits will be legally required. This will create incentives for both AI system and service providers and users to undergo audits.

---

[40] Although it is not directly "AI audit," for example, in "accounting auditing," functional limitations include (1) the relativity of accounting judgments, (2) the limitation of auditor involvement in actual transactions, events, and facts, and (3) the limitation of auditing as a contractual matter. (Yamaura, 2018, p.12)

[41] As mentioned above, New York City, USA, enacted a law in July 2023 regulating AI-based hiring.

**4.2 AI Audit Proof Propositions**

As described in "the Need for AI Audits," they may be conducted based on the accuracy of AI judgment as well as fairness and privacy as proof propositions. What the proof propositions will be depends on the audited company and on societal demands.

**4.3 AI Audit Scope**

This section considers the cases both where an AI system service itself and where the internal controls implemented by the organization providing the AI service are audited.

**4.3.1 Cases where the AI system or service itself is audited**
Examining the items listed in Figure 1, the following, are considered to be audit scope.

**(1) Trained model**

This would include the module files and source code of completed programs as well as the parameter files such as feature values, etc. All relevant documents, such as the design specifications of AI models, are also subject to audit.

**(2) Training Data**

Historical internal personnel data, published statistical data, etc. would be considered for audit scope.

**(3) Software Infrastructure**

The operating system, database, etc., of the server running the AI model or program would be subject to audit. In addition, a server not housed on site but by a cloud service, the relevant software service of which, would also be considered subject to audit.

**(4) Hardware**

The performance of camera equipment, recording equipment, and other devices used to record the applicant interviews that are the AI system's input data would also be subject to audit. In addition, the processing capacity and redundancy of servers running AI models, of any cloud and backup services used, etc., and the regions where they are used, as well as their software infrastructure, are also considered to be subject to audit.

**(5) Other service providing aspects**

The audit would cover whether the use of recruitment AI to make hiring decisions and scoring is disclosed to job seekers, how third-party checks are conducted, etc.

**4.3.2 Cases of auditing the organization and internal controls that provide AI services**

The audit is expected to cover internal controls implemented by (1) management, (2) the human resources department as user division, and (3) the system development and operation department. The following internal control is only one example to be considered. In actual audit cases, the internal control to be considered will be wider ranging, taking into account the audited company's objective thinking, business processes, system development, operational processes, and the use of AI services.

**(1) Management**

To confirm the management's policy for the development and operation of AI services, it is essential to confirm whether the management has established and operated an AI utilization policy and training system, and whether the PDCA cycle is functioning effectively.

**(2) Human Resources Department as user division**

Rules may be established to stipulate that the actual acceptance or rejection decision is made according to the AI's judgment, or, if it determines that a decision cannot be made by AI system, the staff member managing the interview must manually make the decision, requiring confirmation that this is done appropriately. In addition, if such training data is prepared by the Human Resources Department, it is assumed to be a verification procedure for the internal control that assesses whether the criteria for setting the correct labels are defined and whether this label setting's results are checked by multiple persons.

**(3) Systems Department: The development and operations division**

If there are rules for checking and approving various design and test plans and implementation results, audit procedures are assumed to confirm whether their design are appropriate and they are being operated effectively. If there are thresholds for the accuracy of released programs, it may be necessary to confirm that only programs that exceed those thresholds are being released.

### 4.4 Factors that complicate AI audits

When identifying and discussing specific AI services and systems to be audited, issues related to the industry and the context in which they are used make AI audits more complex than discussing them in general terms. For example, regarding the recruitment AI service, additional issues such as employment practices and social systems related to education must be considered, thus complicating the discussion.

#### 4.4.1 Complexity of AI Technology

Even if an appropriate acceptance/rejection judgment is output at the time of the audit, the appropriateness of the acceptance/rejection decision may vary due to continuous learning and business or societal changes when the audit results are used. Expressing the logic of the acceptance/rejection decision as a mathematical formula or conditional judgment could be problematic, and even if the logic can be explicitly confirmed, determining whether it is appropriate could be a challenge.

Also, assumptions must be considered concerning whether the HR department would rely entirely on the recruitment AI's judgment in making hiring decisions. If the AI's judgment is used supplementally and the final decision is made by a person, the audit for recruitment AI would be positioned differently.

#### 4.4.2 Underdevelopment of institutional design for AI audits

It is envisaged that an audit of recruitment AI will be required if the EU AI act comes into force. However, at this stage, no firm general standards have been established, such as requirements for AI auditors and quality management systems, etc. In July 2023, New York City enacted a law regulating employer's use of AI in making hiring decisions. It is expected that this law will serve as a benchmark for future legislation.

#### 4.4.3 Complexity of setting performance standards for AI audits

Indicators for evaluating accuracy and performance must be included when considering criteria for conducting AI audits. For example, it may be assumed that the perspective of bias may not be emphasized in the indicators used to ascertain the accuracy of accurately determining which personnel are excellent and do not retire. Even if the model accurately identifies excellent human resources, there may be cases where discriminatory judgments are made.

Establishing standards for what constitutes a non-biased and what constitutes a fair result, and reaching social consensus thereon is difficult. Although the concept of fair recruitment and selection is already present in the conventional recruitment processes, employers have the discretion to determine their recruitment policies, recruitment criteria, and hiring decisions, and there are no uniform standards to ensure fair recruitment.

Furthermore, while employers can after the fact review whether the AI's judgment was correct, they cannot after the fact review whether the rejected candidate was truly unsuitable, thus limiting the evaluation of accuracy to only some cases.

### 4.4.4 Complexities arising from the scope of the auditee

The issues differ depending on whether the recruitment AI service is being built for a single company or for several organizations. In this case study, it is assumed that the service is developed and operated across the organization.

### (1) Development and operation of AI systems across organizations

When building an AI system, it is difficult to determine the scope of audit coverage when there are outsourcing software companies and their subcontractors which manage actual developing process.

### (2) Use of external libraries

It is a consideration whether or not auditing is required for external libraries employed for machine learning approaches such as deep learning, gradient boosting, random forest, etc. In addition, when models are built using an AI system construction service, the service is presumptively subject to audit. Barriers to auditing these external libraries and construction services will be encountered, and it may be difficult to determine the scope of auditing to be included.

### (3) Management of training data

When data is provided by private recruitment agencies, it must be considered whether or not an audit of their data management system is required. Even if an audit is required, various barriers to conducting an actual audit will be encountered. It is also difficult to determine how public data published by public organizations should be handled. This requires deciding on whether statistical information published by public organizations can be used unconditionally without an audit regardless of the context in which it is utilized.

Furthermore, if the HR department is responsible for correctly labeling the historical personnel data for training the system, the scope of the audit may be extensive because it will be necessary to audit the HR department's labeling processes as well.

### 4.4.5 Balancing the need and supply for AI audits

To a certain extent, recruitment AI is already being used. While the need for auditing high-risk AI, including recruitment AI, may increase in the future, conducting AI audits with appropriate scope and personnel requirements will be difficult due to inconsistent expectations concerning what is assured by the AI audits, the lack of human resources, and lack of incentives.

### (1) Mismatch in expectations regarding what an AI audit assures

As an example, a public expectation could be that claims such as "this AI provides 100% correct answers to the results of acceptance/rejection decisions" or "this AI completely eliminates the biases everyone believes are present", will be audited. Conversely, the limits of what can be confirmed and assured in an actual audit will be limited to confirming whether

"each evaluation indicator is within the predefined accuracy threshold of the AI's judgment parameters regarding acceptance or rejection" and whether "it conforms to the fairness criteria defined by the auditor and management". In such cases, an expectation gap would emerge regarding the content of the AI audit.

**(2) Lack of human resources to conduct AI audits**

Conducting an audit of recruitment AI services requires a wide range of capabilities that includes auditing skills, technical knowledge of AI, and an understanding of the social, legal, and ethical perspectives affecting employment and human resource management processes. In particular, auditors are currently required to design their own procedures and define how AI recruitment services should be audited. There may also be a shortage of personnel with the knowledge and experience to accomplish these tasks.

**(3) Mismatch between supply and demand for AI audit practitioners**

Should a compensation claim arise for unfair discrimination against a job applicant, the potential liability of the AI service's external auditor as well as the management who developed and used the recruitment AI service could become a matter of debate. If such liability is confirmed, auditors will balk at accepting assignments if the audit fees do not match the liability.

**(4) Lack of incentive for companies to undergo AI audits**

By September 2023, there are no laws or regulations in Japan that require the auditing of recruitment AI, so there is little incentive for companies to undergo audits. In addition, if the verification areas covered are wide-ranging, substantial audit fees will be paid to conduct a full-scale external audit. It is conceivable that some companies will stop using recruitment AI because the benefits are not worth the cost of external audits.

**5. Future issues and recommendations on AI auditing**

This paper has summarized the issues to be discussed regarding the auditing of AI services and systems, and the factors that complicate AI audits. Based thereon, this chapter provides recommendations for resolving these issues and promoting AI audits appropriately in the future.

**5.1 Development of institutional design for AI audits**

Currently in Japan, while there is a need to use AI safely and securely, due to lacking institutional design for AI audits, AI auditing has not proliferated. Regarding the international community, the need for institutional design for AI audits is expected to increase, supported by the introduction of the EU AI act. Therefore, we believe that discussions on institutional design for AI audits for Japan, are also necessary. Although the need for auditing AI services and systems has been noted in many policy documents, there remains a wide range of topics such as audit scope, timing, etc., to be discussed before AI auditing is realized. Therefore, discussions should proceed based on common understandings, as presented in this paper, when preparing the institutional design for AI audits and conducting actual audits following them.

In developing institutional design for AI audits by external auditors, the balance between audit quality and fees must also be considered. Audit standards and audit fees that are too low may engender doubts about the quality of the audit results, while high audit standards and fees may become barriers to entry to the introduction of AI services, thereby inhibiting innovation. In terms of audit fees, it is also necessary to consider designing a system that takes into account where audit fees will be borne when multiple organizations are subject to AI audits.

The timing of audits must also be considered in the design of an external AI audit system. In current system audits, new services and systems are often audited after being released. However, the audited company may need to resolve problems prior to the release of the service or system and wish to prepare for a flawless release. Therefore, institutional design for AI audits is needed to meet such pre-release audit needs.

Furthermore, AI services and systems are often developed and used widely across countries and regions, and if AI regulations are developed for each country or region, trade barriers may arise between those requiring audits of AI services and systems and those that do not. From this perspective, it is desirable to promote discussions on AI audits and to consider performance standards that allow for interoperability. The institutional design for AI audits that considers the rapid progress of technology, could be prepared with using the sandbox system.[42]

### 5.2 Training human resources for AI audits

Conducting an AI audit requires diverse expertise with a wide range of skills and experience. In addition to understanding auditing theory, having industry knowledge related to the audited company and the services it provides, and knowledge and experience in Information Technology (IT) areas not limited to AI, AI-specific technical knowledge, ethics and culture, laws and regulations, and a very broad range of other knowledge and experience is required. It must be assumed that, in Japan, the number of human resources possessing such knowledge and experience is still small. The development of auditors capable of responding to future AI auditing needs must be promoted. As explained in Chapter 2, the AI audit relevant parties and organization consists of a wide range of people and organizations. Considering this variety of people involvement, along with professional knowledge and work experience, these tasks require additional development, including soft skills, such as appropriate information sharing with stakeholders and communication skills in a multi-person auditing process. Furthermore, while there are various qualification and certification systems for audit practitioners[43], we believe that a new system of qualification requirements is needed for individual and organizational AI auditors.

### 5.3 Updating AI audits in accordance with technological progress

Technical research related to AI is flourishing around the world, and new services and systems are emerging daily. In addition, the forms of use and methods associated with AI services, as well as organizational and people development, are advancing in tandem with these innovations. Since new risks and challenges are expected to emerge and increase in importance consequent to these changes, the AI audit institutional design, standards and methods must be updated to avoid obsolescence.[44] The update mechanism and the speed and frequency of updates should also be discussed to ensure that they are made proactively as technologies advance while taking into account international interoperability.

### 6. Toward a society that can use AI safely and trustfully

This report is intended to summarize the issues related to AI auditing and to establish a common foundation for discussing critical issues among the parties concerned. This objective has to some extent been met, a discussion must follow on the ideal AI audit and its practical treatment in the future.

---

[42] The Cabinet Secretariat describes the regulatory sandbox system as "a system in which, when it is difficult to commercialize a new technology or implement a new business model in relation to current regulations, a demonstration is conducted based on an application by a business operator, with approval from the regulatory agency, for the social implementation of the new technology or business model, and the information and data obtained through the demonstration are used to review regulations. The information and data obtained from the demonstrations are then used to review the regulations", https://www.cas.go.jp/jp/seisaku/s-portal/regulatorysandbox.html.

[43] For example, in Japan, a certified public accountant qualification as defined by the Institute of Certified Public Accountants of the world, a systems auditor as defined by METI, and a certified internal auditor as defined by IIA, etc.

[44] The "System Audit Standards" (https://www.meti.go.jp/policy/netsecurity/sys-kansa/sys-kansa-2023r.pdf) were established by METI in 1985 and have been revised repeatedly in 1996, 2004, 2018, and 2023.

The current social climate surrounding AI audits includes the European Parliament's adoption on June 14, 2023, of amendments to the EU AI act. [45] This preceded the announcement on July 21, 2023, from the White House that they had secured voluntary commitments that will serve as a guiding principle from leading AI companies to manage the risks posed by AI and to address the broader concerns surrounding AI technology.[46] In addition, a "Committee on AI" set up in the Council of Europe is negotiating to draft the world's first AI treaty. [47]

There is an increasing need to assess and audit the proper governance and management of AI from AI service provider and AI service user companies. Since AI use is expected to advance rapidly, we hope that this paper will contribute to realizing responsible AI development and operation by AI system and service companies, and a society in which AI can be used safely and trustfully.

In addition, the discussion over AI audits has developed rapidly and the knowledge of many experts is indispensable. Since the development of AI technology is also rapid, it is necessary to enhance this policy recommendation in accordance with the latest information and best practices in this dynamic situation. We hope that cooperation with domestic and international experts and relevant organizations will contribute to building more effective and practical AI audit discussions.

## 7. Appendix: AI Auditing and Generative AI

This paper has taken machine learning in general, including deep learning, as an exemplary AI technology and has summarized the issues and difficulties surrounding its audit. Furthermore, generative AI, the use of which has rapidly expanded since 2022, has many of the auditing issues and challenges raised in this paper in common, but ones specific to generative AI are anticipated and introduced here. As a premise, this study group regards generative AI to be that which generates new data, such as text data and image data, by utilizing foundational models, large language models, and other technologies.

## 7.1 The Potential and Challenges of Generative AI

The AI covered by this paper is defined by the OECD as " Machine-based system that can, for a given set of human-defined objectives, make predictions, recommendations or decisions influencing real or virtual environments" and the word "generative" is not included.

Generative AI differs from other AI which performs recognition and prediction by learning from existing content and data, such as images, audio, video, and text, to generate new content and data. However, generative AI is not an entirely novel technology that only emerged after 2022. Rather, it is a type of machine learning method, like conventional AI that performs recognition, prediction, and judgment with its potential and problems having been discussed since the late 2010s. For example, the automatic generation of fake videos of celebrities and works that imitate the style of past greats has been done and identified as a problem. While machine generation can produce rich ideas and expressions that humans could not have devised, associated legal, ethical, social, and economic issues such as copyright infringement, undermining the dignity of creators and those targeted by imitation, fake news, and defamation have become apparent.

---

[45] EU AI Act Compromise text is availabe on European Parliament's website, https://www.europarl.europa.eu/news/en/press-room/20230505IPR84904/ai-act-a-step-closer-to-the-first-rules-on-artificial-intelligence.

[46]More detail about this commitment is available on the White House website, https://www.whitehouse.gov/briefing-room/statements-releases/2023/07/21/fact-sheet-biden-harris-administration-secures-voluntary-commitments-from-leading-artificial-intelligence-companies-to-manage-the-risks-posed-by-ai/.

[47]A consolidated version of the draft became available on July 23, 2023, on Committee on AI of the Council of Europe's website (https://coe.int/en/web/artificial-intelligence/cai) (https://rm.coe.int/cai-2023-18- consolidated-working-draft-framework-convention/1680abde66). For a description of this conference and other information, see the event report of Institute for Future Initiatives, The University of Tokyo and other sources, https://ifi.u-tokyo.ac.jp/project-news/16287/.

What differs between that situation and the situation after 2022? First is that free and fee-based AI services and systems are now available and widely accessible to the general public. Some generative AI services and systems were available before 2022, but access by general users was limited. Since 2022, the number of generative AI users has increased exponentially, and services and systems that use generative AI may, in the future, become clients for AI audits.

**7.2 Issues and Challenges Focused on Generative AI**

Along with AIs for recognition, prediction, etc., generative AI is a one type of machine learning method, so shares the issues and the factors that make AI auditing difficult previously addressed in this paper. Conversely, several issues are unique to or characteristic of generative AI.

For example, in the section on issues surrounding AI auditing in Chapter 2, the proof proposition (2.2) for AI auditing may change. While the perspectives of fairness, transparency, etc. were addressed for recognition and prediction AI, in generation AI and dialogue AI, discussions are active on copyright, false information and misinformation, etc., as well as on the relationship between humans and machines (human-machine interface), including considerations of personal dignity, emotional manipulation, and dependence on AI as a social value or a specific. The adjustment (alignment) of AI to social values and specific domains has been technically verified through methods such as fine-tuning but the development of criteria for properly evaluating value is in its infancy (3.3), which makes auditing generative AI more difficult.

Turning to the burgeoning use of generative AI, in addition to the rapid development of technologies for utilizing generative AI, e.g., by improving the performance of foundation model, large language model and portal, including hardware performance, the widespread use of interactive generative AI are factors that have popularized text and image generative AI since 2022. Therefore, the design of the interface, e.g., whether it is properly explained that confidential information should not be input, etc., could be subject to audit, as discussed in the AI Audit Scope (4.3). Such interfaces increase the complexity of AI technology (3.1) because they are based on the interaction between humans and AI. In the AI audit introduced in this paper, we emphasized the importance of preconditions such as whether the final decision is made by a human or by the collaboration between AI and a human. If the product of the interactive generative AI must be audited, records such as which instructions were given, and the logic controlling the input instructions may also be audited. For example, if the output from the generated AI is controlled by excluding or correcting inappropriate input instructions, the confirmation for such exclusion or correction logic is working properly assumed to be important part of audit procedure. In addition, when auditing the output of generative AI, an audit of whether appropriate instructions are provided for using the generated output is also expected to be an audit procedure for generative AI. The EU AI act requires that the output content of generative AI be marked as having been generated by AI. One of the issues to be discussed is whether the outputs from generative AI comply these requirements, similar to watermarking.

In most cases, the functionality of generative AI is achieved by individual AI systems referring to a large-scale model that serves as the core of the system. For example, a large language model is applicable for a sentence generation AI. In addition, various tools that support the use of individual AI systems are being developed all the time. In the case of generative AI, there are many systems, tools, and parties involved in the use of an AI service or system, and in many cases, they cross national borders and organizations, making it more difficult than ever to understand all the parties involved.

Since, in light of the above, there are some issues which will be focused on generative AI, it is important to continually update the issues and methods of auditing as technology develops in the future.

## References

**In Japanese**
Akoshima, T., & Fukuda, S. 2020a. Considering Internal Audit of AI System (No.1) Trend and Issues of AI Application. Junkan Keirijoho, *1592*, 50-54.
Akoshima, T., & Fukuda, S. 2020b. Considering the internal audit of AI system (No.2) Overall view of AI system audit. Junkan Keirijoho, *1595*, 56-60.
Akoshima, T., & Fukuda, S. 2020c. Considering the Internal Audit of AI System (No.3, Completed) Implementation Model of AI System Audit. Junkan Keirijoho, *1596*, 46-52.
Toba, Yoshihide, "Audit of Financial Statements: Theory and System [Basics]," Kokugen Shobo, 2009.
Hisashi Yamaura, "Accounting and Auditing Theory," Chuokeizai-sha, 2008.
IPA, "Common Frame 2013," IPA, 2013.
Ichiro Akimoto, "AI Business Compendium," PRESIDENT Inc, 2022.

**In English**
Bandy, J. 2021. "Problematic Machine Behavior: A Systematic Literature Review of Algorithm Audits", Proceedings of the ACM on Human-Computer Interaction, Vol.5, No.CSCW1,Article 74.pp.1-34.
de Boer, M. 2023. "Trustworthy AI and accountability: yes, but how? What the EU AI Act's approach to AI accountability can learn from the science of algorithm audit". [Thesis, externally prepared, Universiteit van Amsterdam].
Batarseh, F. A., Freeman, L., and Huang, C.-H. 2021. "A survey on artificial intelligence assurance", Journal of Big Data, Vol.8, 60.
Dwork, C. 2012. "Fairness Through Awareness", Proc. of the 3rd Innovations in Theoretical Computer Science Conf: 214-226.
Hardt, M. 2016. "Equality of Opportunity in Supervised Learning", Advances in Neural Information Processing Systems 29: 3323-3331.
Institution of Internal Auditors (IIA), 2017a. "Artificial Intelligence – Considerations for the Profession of Internal Auditing," Global Perspective and Insights (Special Edition).
IIA, 2017b. "The IIA's Artificial Intelligence Auditing Framework Practical Applications, Part A," Global Perspective and Insights (Special Edition),
IIA, 2017c. "The IIA's Artificial Intelligence Auditing Framework Practical Applications, Part B," Global Perspective and Insights (Special Edition).
IIA, 2017d. "International Standards for the Professional Practice of Internal Auditing (IPPF)"
IIA, 2020. "The IIA's Three Lines Model: An update of the Three Lines of Defense"(IIA Position Paper).
Issa, H., T. Sun, M. Vasarhelyi, 2016. "Research Ideas for Artificial Intelligence in Auditing: The Formalization of Audit and Workforce Supplementation," Journal of Emerging Technologies in Accounting, Vol. 13, No. 2, pp.1-20.
Jobin,A., Ienca, M., Vayena, E. 2019. "The global landscape of AI ethics guidelines," Nature, Machine Intelligence, 1, 389-99.
Koshiyama, A., E. Kazim and P. Treleaven, 2022. "Algorithm Auditing: Managing the Legal, Ethical, and Technological Risks of Artificial Intelligence, Machine Learning, and Associated Algorithms", Computer, Vol.55, No.4, pp.40-50.
Mökander, J., J. Schuett, H. Kirk, L. Floridi, 2023, "Auditing Large Language Models: A Three-Layered Approach," arXiv:2302.08500.
Nakano, M., 2023. "Toward Building the Framework for Artificial Intelligence Audit Theory", Industrial Technology No.45, pp.52-55.
New York City Local Law 144, 2021.

Acknowledgements

This research was conducted as part of the University of Tokyo's Institute for Future Initiative's flagship project "Designing Future Vision in AI Society" as well as research under the "University of Tokyo Distinguished Researcher" program. In addition, the results of joint research with companies and government officials in the AI Governance Project are also included as part of this recommendation.

Many people provided valuable advice in the writing of this policy recommendation. Due to time constraints and organizational affiliations, we are unable to include all their names, but acknowledge and thank them for the useful feedback.

Takashi Akoshima (Member of Japan Society for Systems Audits, Certified Internal Auditor)
Naoto Ichihara (Ernst & Young ShinNihon LLC, AI Leader)
Koichi Ito (PricewaterhouseCoopers Aarata LLC Assurance Innovation & Technology Partner / Deputy Chief of AI Audit Lab)
Teppei Usuki (KPMG AZSA LLC, Digital Innovation, Partner)
Nobuhiko Kato (Ernst & Young ShinNihon LLC, Digital Leader)
Yuji Shimada (Guest Researcher, Research Institute of Industrial Technology, Toyo University)
Kiriko Shimizu (PricewaterhouseCoopers Aarata LLC Assurance Innovation & Technology Senior Manager)
Hideaki Shiroyama (Professor, Institute for Future Initiatives, The University of Tokyo)
Hiroshi Taki (Professor (Professor of Auditing), College of Business Administration, Ritsumeikan University)
Hiroshi Naka (Professor, Institute for Future Initiatives, The University of Tokyo / 2023–24 Director-Research for the Board of Trustees (BOT) on the Global Board of Directors of The Institute of Internal Auditors (IIA))